\documentclass[12pt]{article}
\usepackage{epsfig}
\usepackage{cite}

\def\Title#1{\begin{center} {\Large {\bf #1} } \end{center}}

\newcommand{\RTau}[1]{R_{\tau, \mbox{\tiny #1}}}
\newcommand{\DeltaQCD}[1]{\Delta^{\mbox{\tiny #1}}_{\mbox{\tiny QCD}}}
\newcommand{\ind}[2]{^{\mbox{\scriptsize $#1$}}_{\mbox{\scriptsize #2}}}
\def\Vud{V_{\mbox{\scriptsize ud}}}
\def\Sew{S_{\mbox{\tiny EW}}}
\def\dpew{\delta'_{\mbox{\tiny EW}}}
\def\MTau{M_{\tau}}
\def\Nc{N_{\mbox{\scriptsize c}}}
\def\nf{n_{\mbox{\scriptsize f}}}

\def\va{_{\mbox{\tiny V/A}}}
\def\zva{\zeta\va}

\begin{document}

\Title{The effects due to hadronization \\[1.5mm]
in the inclusive $\tau$~lepton decay}

\bigskip\bigskip


\begin{raggedright}

{\it A.V.~Nesterenko\index{Nesterenko, A.V.}\\
Bogoliubov Laboratory of Theoretical Physics \\
Joint Institute for Nuclear Research \\
Dubna, 141980, Russian Federation \\
E--mail: nesterav@theor.jinr.ru}
\bigskip\bigskip
\end{raggedright}

\section{Introduction}
\label{Sect:Intro}

The main subject of this paper is the role of effects due to hadronization
in the theoretical description of inclusive $\tau$~lepton decay. This
paper employs the results of the studies of this strong interaction
process elaborated over past several years~\cite{PRD64, MAPT1, MAPT2,
NPQCD07, MAPT3, MAPT4} and presents some new results recently obtained in
this field~\cite{Prep}.

The $\tau$~lepton is the only lepton which is heavy enough to decay into
hadrons. This feature enables one to use this process in tests of Quantum
Chromodynamics~(QCD) and entire Standard Model. The theoretical
description of inclusive $\tau$~lepton hadronic decay, similarly to the
case of electron--positron annihilation into hadrons, requires no such
phenomenological models as, for example, the so--called ``Parton
Distribution Functions'' involved in the analysis of Deep Inelastic
Scattering processes. It is worthwhile to note also that the experimental
measurements of $\tau$~lepton decay are of a high accuracy. But the most
interesting feature of this process is that it probes the hadron dynamics
at energies below the mass of $\tau$~lepton.

The experimentally measurable quantity here is the ratio of the total
width of $\tau$~lepton decay into hadrons to the width of its leptonic
decay, which can be decomposed into three parts, namely
\begin{equation}
\label{RTauGen}
R_{\tau} = \frac{\Gamma(\tau^{-} \to \mbox{hadrons}^{-}\, \nu_{\tau})}
{\Gamma(\tau^{-} \to e^{-}\, \bar\nu_{e}\, \nu_{\tau})} =
\RTau{V} + \RTau{A} + \RTau{S}.
\end{equation}
In the right--hand side of this equation, the first two terms account for
the hadronic decay modes involving light quarks (u,~d) only and associated
with vector~(V) and axial--vector (A) quark currents, respectively,
whereas the last term accounts for the decay modes which involve strange
quark. Each of the first two terms can be further decomposed into two
parts according to the angular momentum in the hadronic rest frame, namely
\begin{equation}
\label{RTauGen2}
\RTau{V} = \RTau{V}^{\mbox{\tiny $J$=0}} + \RTau{V}^{\mbox{\tiny $J$=1}},
\qquad
\RTau{A} = \RTau{A}^{\mbox{\tiny $J$=0}} + \RTau{A}^{\mbox{\tiny $J$=1}}.
\end{equation}
In what follows we shall restrict ourselves to the consideration of
parts~$\RTau{V}^{\mbox{\tiny $J$=1}}$ and~$\RTau{A}^{\mbox{\tiny $J$=1}}$
of ratio $R_{\tau}$~(\ref{RTauGen}).

\section{Theoretical description of $\tau$~lepton decay}
\label{Sect:TauTheor}

The theoretical prediction for the quantities on hand~(\ref{RTauGen2})
reads
\begin{equation}
\RTau{V/A}^{\mbox{\tiny $J$=1}} = \frac{\Nc}{2}\,|\Vud|^2\,\Sew
\Bigl(\DeltaQCD{V/A} + \dpew \Bigr),
\end{equation}
where $\Nc=3$ is the number of colors, $|\Vud| = 0.9738 \pm 0.0005$ is
Cabibbo--Kobayashi--Maskawa matrix element~\cite{PDG2010}, $\Sew = 1.0194
\pm 0.0050$ and $\dpew = 0.0010$ stand for the electroweak corrections
(see Refs.~\cite{BNP, EWF1, EWF2}), and
\begin{equation}
\label{DeltaQCDDef}
\DeltaQCD{V/A} = 2\int_{m\va^2}^{\MTau^2}\!
f\!\biggl(\frac{s}{\MTau^2}\biggr) R^{\mbox{\tiny V/A}}(s)\,
\frac{d s}{\MTau^2}
\end{equation}
denotes the QCD contribution. Here $\MTau=1.777\,$GeV is the mass of
$\tau$~lepton~\cite{PDG2010}, $m\va$~stands for the total mass of the
lightest allowed hadronic decay mode of $\tau$~lepton in the corresponding
channel, $f(x) = (1-x)^{2}\,(1+2x)$, and
\begin{equation}
\label{RDef}
R^{\mbox{\tiny V/A}}(s) = \frac{1}{2 \pi i}
\lim_{\varepsilon \to 0_{+}}
\Bigl[\Pi^{\mbox{\tiny V/A}}(s + i \varepsilon) -
\Pi^{\mbox{\tiny V/A}}(s - i \varepsilon)\Bigr]\! =
\frac{1}{\pi}\,\mbox{\normalfont Im}\!\lim_{\varepsilon \to 0_{+}}
\!\Pi^{\mbox{\tiny V/A}}(s + i \varepsilon),
\end{equation}
with~$\Pi^{\mbox{\tiny V/A}}(q^2)$ being the hadronic vacuum polarization
function. In what follows the superscripts~``V'' and~``A'' will only be
shown when relevant.

In general, it is convenient to perform the theoretical analysis of
inclusive $\tau$~lepton decay in terms of the Adler function~\cite{Adler}
\begin{equation}
\label{AdlerDef}
D(Q^2) = - \frac{d\, \Pi(-Q^2)}{d \ln Q^2}, \qquad
Q^2 =-q^2=-s.
\end{equation}
In the framework of perturbation theory its ultraviolet behavior can be
approximated by power series in the strong running
coupling~$\alpha\ind{}{s}(Q^2)$
\begin{equation}
\label{AdlerPert}
D(Q^2) \simeq  D\ind{(\ell)}{pert}(Q^2) = 1 + \sum\nolimits_{j=1}^{\ell}
d_{j} \Bigl[\alpha\ind{(\ell)}{s}(Q^2)\Bigr]^{j},
\qquad Q^2\to\infty,
\end{equation}
where at the one--loop level (i.e., for $\ell=1$)~$\alpha\ind{(1)}{s}(Q^2)
= 4\pi/(\beta_{0}\,\ln z)$, $z=Q^2/\Lambda^2$, $\beta_{0}=11-2\nf/3$,
$\Lambda$~denotes the QCD scale parameter, $\nf$~is the number of active
flavors ($\nf=2$ will be assumed hereinafter), and~$d_{1}=1/\pi$, see
papers~\cite{AdlerPert3La, AdlerPert3Lb, AdlerPert4LPrelim} and references
therein for the details. It is worth noting also that the
function~$R(s)$~(\ref{RDef}) and the Adler function~(\ref{AdlerDef}) can
be expressed in terms of each other by making use of the following
relations (see Refs.~\cite{Rad82, KP82, Adler} for the details)
\begin{equation}
R(s) = \frac{1}{2 \pi i} \lim_{\varepsilon \to 0_{+}}
\int_{s + i \varepsilon}^{s - i \varepsilon}\!
D(-\zeta)\,\frac{d \zeta}{\zeta}
\qquad \longleftrightarrow \qquad
\label{AdlerDisp}
D(Q^2) = Q^2\! \int_{m^2}^{\infty}\!
\frac{R(s)}{(s + Q^2)^2} \, d s\, .
\end{equation}
In first of these equations the integration contour in the complex
$\zeta$--plane lies in the region of analyticity of the integrand.

It is important to outline here that all the mentioned above is only valid
for ``true physical'' hadronic vacuum polarization
function~$\Pi\ind{}{phys}(q^2)$ and Adler function~$D\ind{}{phys}(Q^2)$.
However, as it often happens, one has to deal with their perturbative
approximations~$\Pi\ind{}{pert}(q^2)$ and~$D\ind{}{pert}(Q^2)$, which are
valid in the ultraviolet asymptotic only. Besides,
expressions~$\Pi\ind{}{pert}(q^2)$ and~$D\ind{}{pert}(Q^2)$ are
inconsistent with dispersion relation~(\ref{AdlerDisp}), which is
determined by the kinematics of physical process on hand.

Thus, one arrives at the point where the results of perturbation theory
need to be ``merged'' with relevant dispersion relations. This objective
can be achieved in the framework of ``Dispersive approach'' to~QCD, which
will be briefly overviewed in Sect.~\ref{Sect:TauDisp}. The theoretical
description of inclusive $\tau$~lepton hadronic decay within Dispersive
approach will be performed in Sect.~\ref{Sect:TauDisp}, whereas the
analysis of this process within perturbative approach will be discussed in
Sect.~\ref{Sect:TauPert}.

\section{Perturbative approach}
\label{Sect:TauPert}

In this Section, we shall study the massless limit, that implies that the
masses of all final state particles are neglected. In this case, by making
use of definitions~(\ref{RDef}) and~(\ref{AdlerDef}), integrating by
parts, and employing Cauchy theorem, the
quantity~$\DeltaQCD{}$~(\ref{DeltaQCDDef}) can be represented as
\begin{equation}
\label{DeltaQCDCauchy}
\DeltaQCD{} = \frac{1}{2\pi}
\int_{-\pi}^{\pi} D\Bigl(M_{\tau}^{2}\,e^{i\theta}\Bigr)
\!\Bigl(1 + 2e^{i\theta} - 2e^{i3\theta} -e^{i4\theta}\Bigr)\, d \theta,
\end{equation}
see,~e.g., papers~\cite{BNP, DP1, DP2, ALEPH06} and references therein. In
general, in Eqs.~(\ref{RDef}) and~(\ref{AdlerDef}) it is convenient to
handle the leading contributions (i.e., the terms of $0$-th~order in the
strong running coupling) separately from the contributions due to strong
interaction, namely
\begin{equation}
R(s)   = r^{(0)}(s)   + r^{(\ell)}(s),
\qquad
D(Q^2) = d^{(0)}(Q^2) + d^{(\ell)}(Q^2).
\end{equation}
In what follows we shall restrict ourselves to the one--loop
level~($\ell=1$).

In fact, the only available option within perturbative approach is to
directly use in the theoretical expression for~$\DeltaQCD{}$ (despite of
remarks given in Sect.~\ref{Sect:TauTheor}) the perturbative approximation
of hadronic vacuum polarization function~$\Pi\ind{}{pert}(q^2)$ instead of
its unknown ``true physical'' expression~$\Pi\ind{}{phys}(q^2)$. For the
case of functions~(\ref{DeltaQCDCauchy}) and~(\ref{AdlerPert}), this
prescription eventually results in (see Ref.~\cite{Prep} for the details)
\begin{equation}
\label{AdlerPMP0}
r^{(0)}(s) = 1
\qquad \longleftrightarrow \qquad
d^{(0)}(Q^2) = 1,
\end{equation}
\begin{equation}
\label{DeltaQCDPert}
\Delta\ind{}{pert} = 1 + \frac{4}{\beta_{0}}\int_{0}^{\pi}
\frac{c_0 A_{1}(\theta)+\theta A_{2}(\theta)}{\pi(c_{0}^{2}+\theta^2)}
\,d\theta,
\end{equation}
where~$c_0 = \ln \bigl( \MTau^2/\Lambda^2 \bigr)$ and
\begin{equation}
A_{1}(\theta) = 1 + 2\cos(\theta) - 2\cos(3\theta) - \cos(4\theta),
\quad\,
A_{2}(\theta) = 2\sin(\theta) - 2\sin(3\theta) - \sin(4\theta).
\end{equation}

\begin{figure}[t]
\begin{center}
\epsfig{file=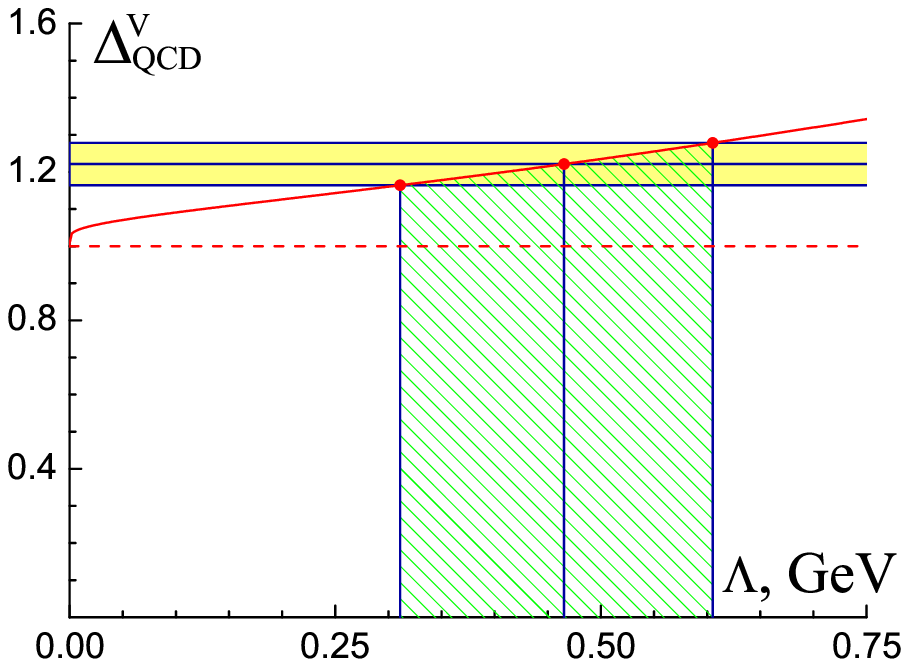,width=70mm}
\hfill
\epsfig{file=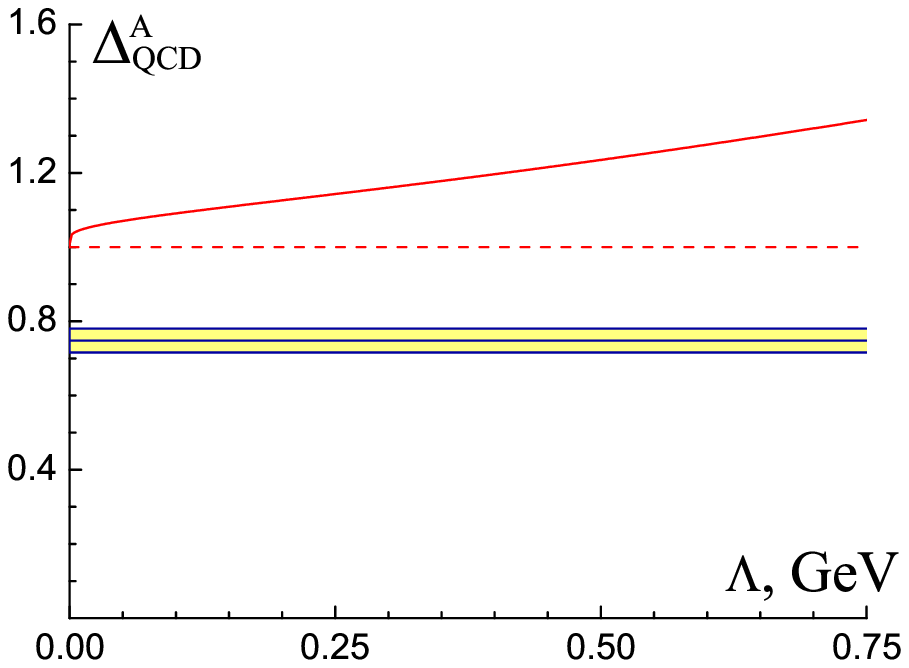,width=70mm}
\caption{Comparison of the one--loop perturbative expression
$\Delta\ind{}{pert}$~(\ref{DeltaQCDPert}) (solid curves) with relevant
experimental data~(\ref{DeltaQCDExp}) (horizontal shaded bands). The
leading--order terms of~$\Delta\ind{}{pert}$~(\ref{DeltaQCDPert}) are
denoted by horizontal dashed lines. The solution for QCD scale
parameter~$\Lambda$ (if exists) is shown by vertical dashed band. The
left and right plots correspond to vector and axial--vector channels,
respectively.}
\label{Plot:Pert}
\end{center}
\end{figure}

Let us proceed now to the comparison of one--loop perturbative
result~(\ref{DeltaQCDPert}) with corresponding experimental data. First of
all, it is worth emphasizing here that perturbative approach gives
identical predictions for functions~$\DeltaQCD{V/A}$~(\ref{DeltaQCDDef})
in vector and axial--vector channels (i.e., $\Delta\ind{\mbox{\tiny
V}}{pert} \equiv \Delta\ind{\mbox{\tiny A}}{pert}$). However, their
experimental values extracted from data presented in Refs.~\cite{ALEPH98,
ALEPH05, ALEPH06} are different, namely
\begin{equation}
\label{DeltaQCDExp}
\Delta\ind{\mbox{\tiny V}}{exp} = 1.221 \pm 0.057, \qquad
\Delta\ind{\mbox{\tiny A}}{exp} = 0.748 \pm 0.032.
\end{equation}
These quantities are juxtaposed with perturbative
result~(\ref{DeltaQCDPert}) in Fig.~\ref{Plot:Pert}. As one can infer from
this figure, for vector channel the corresponding value of QCD scale
parameter is $\Lambda=\bigl(465^{+140}_{-154}\bigr)\,$MeV (formally, there
is also the second solution, $\Lambda=\bigl(1646^{+26}_{-29}\bigr)\,$MeV,
which will not be considered hereinafter). As for the axial--vector
channel, the perturbative approach fails to describe experimental data on
$\tau$~lepton hadronic decay, since for any value of~$\Lambda$ the
function $\Delta\ind{}{pert}$~(\ref{DeltaQCDPert}) exceeds
$\Delta\ind{\mbox{\tiny A}}{exp}$~(\ref{DeltaQCDExp}).

\section{Dispersive approach}
\label{Sect:TauDisp}

\subsection{General remarks}

It is crucial to emphasize that the analysis presented in
Sect.~\ref{Sect:TauPert} entirely leaves out the effects due to
hadronization, which play an important role in the studies of strong
interaction processes at low energies. Specifically, the mathematical
realization of the physical fact, that in a strong interaction process no
final state hadrons can be produced at energies below the total mass of
the lightest allowed hadronic final state, consists in the fact that the
beginning of cut of corresponding hadronic vacuum polarization
function~$\Pi(q^2)$ in complex $q^2$--plane is located at the threshold of
hadronic production, but not at the point~$q^2=0$. Such limitations are
inherently embodied within relevant dispersion relations, which, in turn,
impose stringent physical nonperturbative constraints on the quantities on
hand. Obviously, these restrictions should certainly be accounted for when
one is trying to go beyond the limits of perturbation theory.

The nonperturbative constraints, which dispersion
relation~(\ref{AdlerDisp}) imposes on the Adler function~(\ref{AdlerDef}),
have been merged with perturbative result~(\ref{AdlerPert}) in the
framework of Dispersive approach to QCD, that has eventually led to the
following integral representations for functions~(\ref{RDef})
and~(\ref{AdlerDef}) (see Refs.~\cite{MAPT2, NPQCD07, MAPT4}
for the details):
\begin{eqnarray}
\label{RMAPT}
R(s) &=& r\ind{(0)}{}(s) + \theta\biggl(\!1-\frac{m^2}{s}\biggr)\!
\!\int_{s}^{\infty}
\rho(\sigma)\, \frac{d \sigma}{\sigma}\, , \\[2.5mm]
\label{AdlerMAPT}
D(Q^2) &=& d\ind{(0)}{}(Q^2) + \frac{Q^2}{Q^2+m^2}
\int_{m^2}^{\infty} \rho(\sigma)\,
\frac{\sigma - m^2}{\sigma+Q^2}\,
\frac{d \sigma}{\sigma}\, ,
\end{eqnarray}
where $\theta(x)$ is the unit step--function ($\theta(x)=1$ if $x \ge 0$
and $\theta(x)=0$ otherwise) and~$\rho(\sigma)$ denotes the so--called
spectral density. It is worth mentioning that in the massless limit ($m
\to 0$) expressions~(\ref{RMAPT}) and~(\ref{AdlerMAPT}) become identical
to those of the so--called ``Analytic perturbation theory''~\cite{APT1,
APT2, APT3, APT4}. But it is essential to keep the hadronic mass~$m$
nonvanishing within the approach on hand.

Let us proceed now to the description of inclusive $\tau$~lepton hadronic
decay within Dispersive approach. It is worthwhile to note here that there
are two distinctions between the approach on hand and the massless
perturbative approach presented in~Sect.~\ref{Sect:TauPert}. Specifically,
the first distinction is the incorporation of effects due to
hadronization, and the second one is the expression for the one--loop
spectral density
\begin{equation}
\label{RhoDef}
\rho(\sigma) = \frac{4}{\beta_{0}}\frac{1}{\ln^{2}(\sigma/\Lambda^2)+\pi^2} +
\frac{\Lambda^2}{\sigma},
\end{equation}
which resembles previously studied nonperturbative model~\cite{PRD64,
MAPT1, MAPT4, PRD62, Review, NPBPS04}. In the right--hand side of
Eq.~(\ref{RhoDef}) the first term is the one--loop perturbative
contribution whereas the second term represents intrinsically
nonperturbative part of the spectral density.

\subsection{Abrupt kinematic threshold}

\begin{figure}[t]
\begin{center}
\epsfig{file=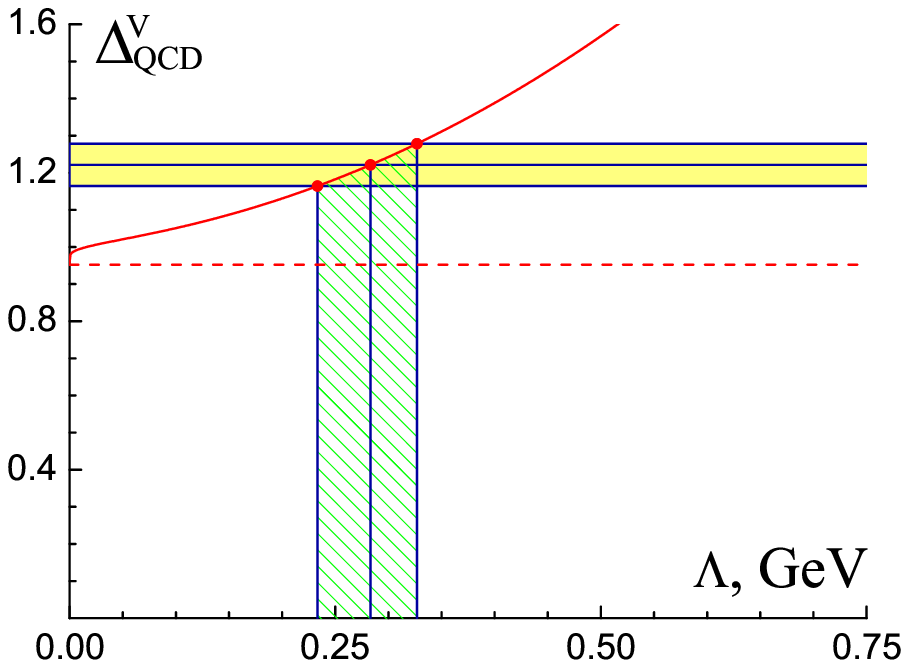,width=70mm}
\hfill
\epsfig{file=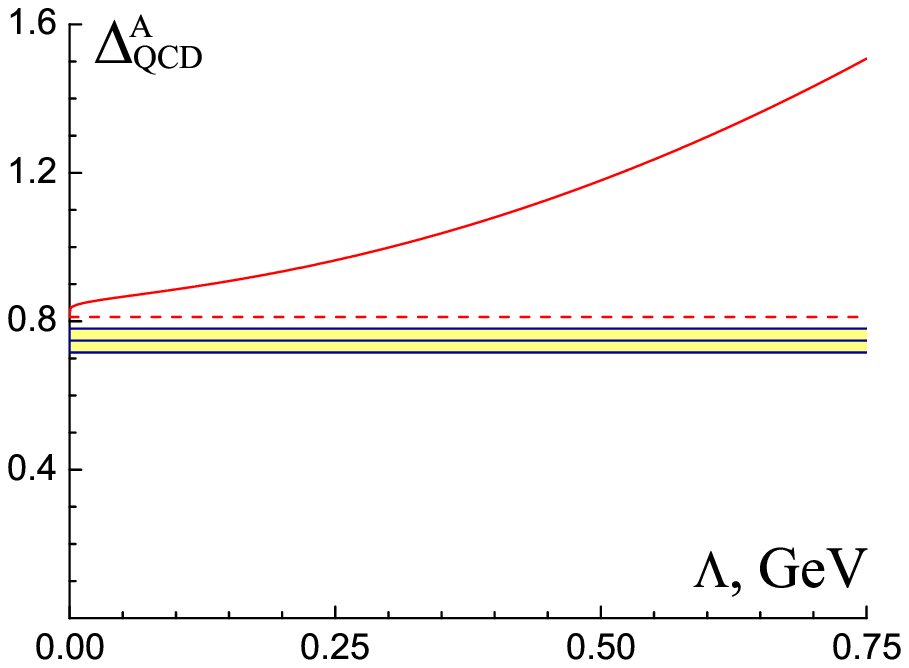,width=70mm}
\caption{Comparison of expression~(\ref{DeltaQCD_MAPT_AT}) (solid curves)
with relevant experimental data~(\ref{DeltaQCDExp}) (horizontal shaded
bands). The leading--order terms
of~$\DeltaQCD{V/A}$~(\ref{DeltaQCD_MAPT_AT}) are denoted by horizontal
dashed lines. The solution for QCD scale parameter~$\Lambda$ (if~exists)
is shown by vertical dashed band. The left and right plots correspond to
vector and axial--vector channels, respectively.}
\label{Plot:MAPT_AT}
\end{center}
\end{figure}

There are two options in the framework of Dispersive approach. The first
one is the so--called ``abrupt kinematic threshold'', which implies that
the leading--order term of~$R(s)$~(\ref{RDef}) is approximated by the
step--function
\begin{equation}
\label{LTAKT}
r\ind{(0)}{\mbox{\tiny V/A}}(s) = \theta\biggl(\!1-\frac{m\va^{2}}{s}\!\biggr)
\qquad \longleftrightarrow \qquad
d\ind{(0)}{\mbox{\tiny V/A}}(Q^2) = \frac{Q^2}{Q^2+m\va^{2}}\, ,
\end{equation}
see papers~\cite{MAPT2, NPQCD07, MAPT4} and references therein for the
details. Equation~(\ref{LTAKT}) accounts only for basic kinematic
restriction on hadronic vacuum polarization function~$\Pi(q^2)$ and, in
fact, represents a rather rough approximation. In this case the quantity
$\DeltaQCD{V/A}$~(\ref{DeltaQCDDef}) reads~\cite{MAPT4}
\begin{equation}
\label{DeltaQCD_MAPT_AT}
\DeltaQCD{V/A} = 1 - g\bigl(\zeta\va\bigr) +
\int_{m\va^{2}}^{\infty}\!H\!\biggl(\frac{\sigma}{M_{\tau}^{2}}\biggr)
\rho(\sigma)\,\frac{d \sigma}{\sigma},
\end{equation}
where $\zeta\va = m\va^{2}/\MTau^{2}$ and
\begin{equation}
H(x) = g(x)\,\theta(1-x) + g(1)\,\theta(x-1) - g\bigl(\zeta\va\bigr),
\qquad
g(x) = x (2 - 2x^2 + x^3).
\end{equation}
However, similarly to perturbative case~(\ref{DeltaQCDPert}),
expression~(\ref{DeltaQCD_MAPT_AT}) is unable to describe the experimental
data on $\tau$~lepton hadronic decay in axial--vector
channel~(\ref{DeltaQCDExp}), see Fig.~\ref{Plot:MAPT_AT}.

\subsection{Smooth kinematic threshold}

The second, more accurate, option within Dispersive approach is the
so--called ``smooth kinematic threshold''. In this case the leading--order
term of $R(s)$~(\ref{RDef}) takes the following form~\cite{Prep}
\begin{equation}
r\ind{(0)}{\mbox{\tiny V/A}}(s) =
\biggl(\!1-\frac{m\va^{2}}{s}\!\biggr)^{{\!}3/2}
\;\;\; \longleftrightarrow \;\;\;\:
d\ind{(0)}{\mbox{\tiny V/A}}(Q^2) = 1 + \frac{3}{\xi}
\biggl\{\!1 + u(\xi)\ln\!\sqrt{1 \!+\! 2\xi\bigl[1 \!-\! u(\xi)\bigr]}\biggr\},
\end{equation}
where $u(\xi)=\sqrt{\strut 1+\xi^{-1}}$ and~$\xi=Q^2/m\va^{2}$. Eventually this
leads to the following expression for the quantity
$\DeltaQCD{V/A}$~(\ref{DeltaQCDDef}):
\begin{eqnarray}
\label{DeltaQCD_MAPT_ST}
\DeltaQCD{V/A} &=& \sqrt{1-\zva}\,\biggl(1 + 6\zva - \frac{5}{8}\zva^{2}
+\frac{3}{16}\zva^{3}\biggr)
+ \int_{m\va^{2}}^{\infty}\!H\!\biggl(\frac{\sigma}{M_{\tau}^{2}}\biggr)
\rho(\sigma)\frac{d \sigma}{\sigma} \nonumber \\
&&-3\zva \biggl(1 + \frac{1}{8}\zva^{2} - \frac{1}{32}\zva^{3} \biggr)
\ln\biggl[\frac{2}{\zva}\Bigl(1+\sqrt{1-\zva}\Bigr)-1\biggr],
\end{eqnarray}
see paper~\cite{Prep} and references therein for the details.

\begin{figure}[t]
\begin{center}
\epsfig{file=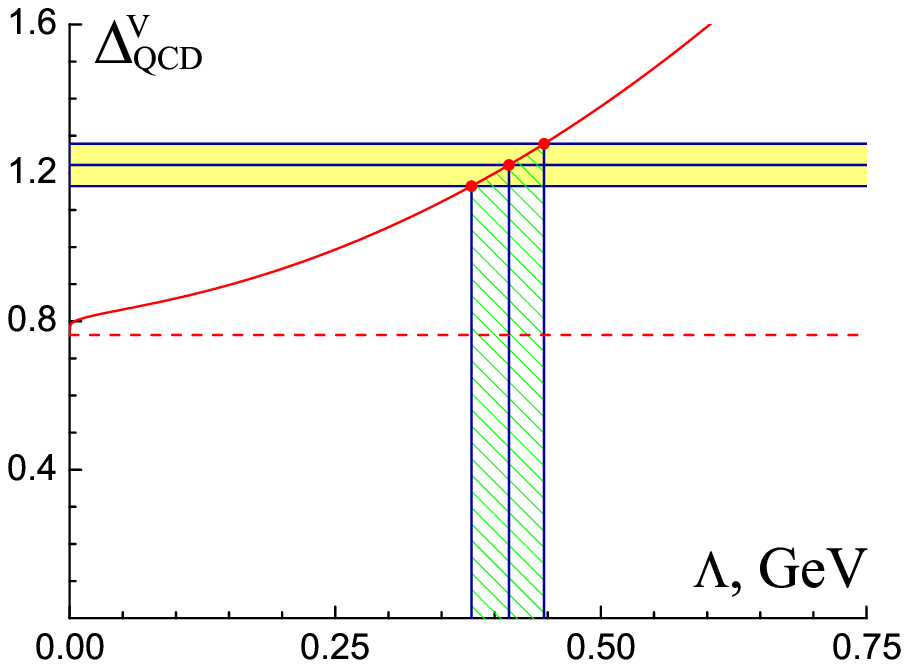,width=70mm}
\hfill
\epsfig{file=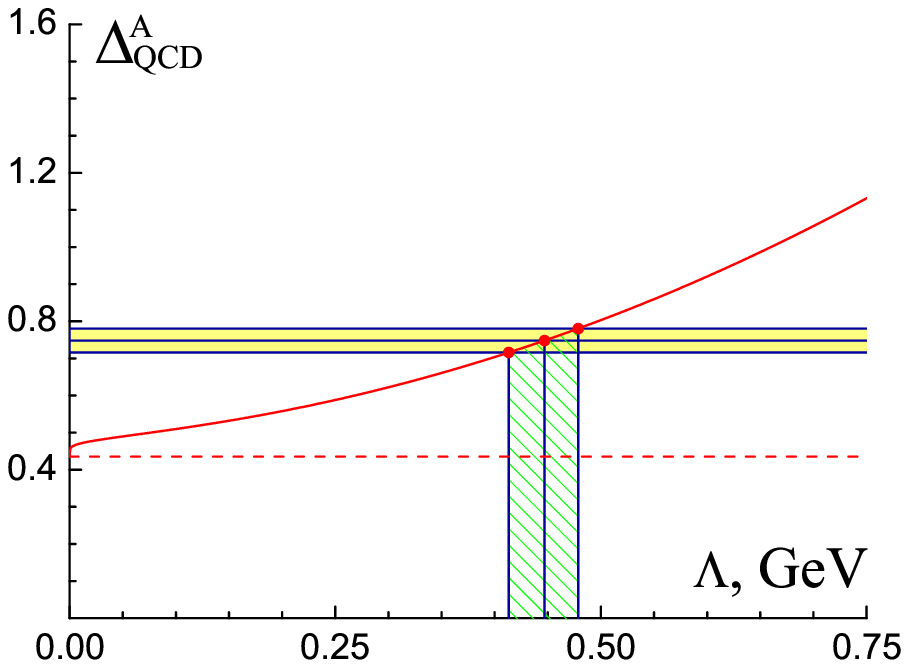,width=70mm}
\caption{Comparison of expression~(\ref{DeltaQCD_MAPT_ST}) (solid curves)
with relevant experimental data~(\ref{DeltaQCDExp}) (horizontal shaded
bands). The leading--order terms
of~$\DeltaQCD{V/A}$~(\ref{DeltaQCD_MAPT_ST}) are denoted by horizontal
dashed lines. The solutions for QCD scale parameter~$\Lambda$ are shown by
vertical dashed bands. The left and right plots correspond to vector and
axial--vector channels, respectively.}
\label{Plot:MAPT_ST}
\end{center}
\end{figure}

It is worth noting also that in the massless limit ($m \to 0$) both
equations~(\ref{DeltaQCD_MAPT_AT}) and~(\ref{DeltaQCD_MAPT_ST})
acquire the same form
\begin{equation}
\label{DeltaQCD_APT}
\DeltaQCD{} = 1 + \int_{0}^{\infty}\!h\biggl(\frac{\sigma}{M_{\tau}^{2}}\biggr)
\rho(\sigma)\,\frac{d \sigma}{\sigma}, \qquad
h(x) = g(x)\,\theta(1-x) + g(1)\,\theta(x-1).
\end{equation}
In the perturbative case the difference between
expressions~(\ref{DeltaQCD_APT}) and~(\ref{DeltaQCDPert})
is due to the residue term
\begin{equation}
\Delta\ind{}{res}\!=\!\frac{4}{\beta_{0}}\,
h_{1}\!\biggl(\frac{\Lambda^2}{M_{\tau}^{2}}\biggr), \quad
h_{1}(x)\!=\!h_{2}(x)\theta(1-x) + h_{2}(1)\theta(x-1), \quad
h_{2}(x)\!=\! x(2-2x^2-x^3),
\end{equation}
which appears to be additionally accounted for in
Eq.~(\ref{DeltaQCDPert}), see discussion of this issue in
paper~\cite{Prep}.

\begin{figure}[t]
\begin{center}
\epsfig{file=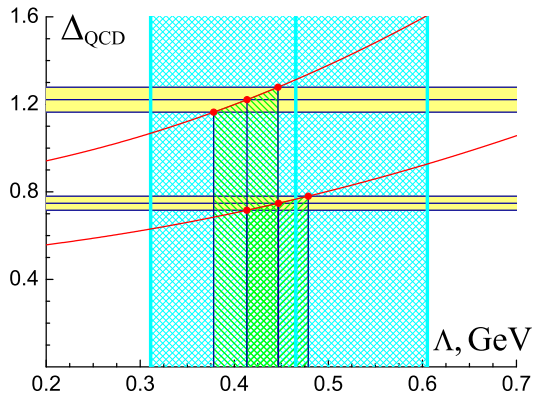,width=70mm}
\caption{The solutions for QCD scale parameter~$\Lambda$ obtained within
Dispersive approach~(\ref{DeltaQCD_MAPT_ST}) for vector and axial--vector
channels (vertical dashed green bands) and perturbative
solution~(\ref{DeltaQCDPert}) corresponding to vector channel (vertical
dashed light--blue band).}
\label{Plot:MAPT}
\end{center}
\end{figure}

The comparison of obtained result~(\ref{DeltaQCD_MAPT_ST}) with
experimental data~(\ref{DeltaQCDExp}) yields nearly identical solutions
for QCD scale parameter~$\Lambda$ in both channels, see
Fig.~\ref{Plot:MAPT_ST}. Namely, $\Lambda = \bigl(412 \pm 34 \bigr)\,$MeV
for vector channel and $\Lambda = \bigl(446 \pm 33 \bigr)\,$MeV for
axial--vector one. Besides, as one can infer from Fig.~\ref{Plot:MAPT},
both these solutions for QCD scale parameter agree very well with
perturbative solution $\Lambda=\bigl(465^{+140}_{-154}\bigr)\,$MeV
obtained in Sect.~\ref{Sect:TauPert}.

\section{Conclusions}

Theoretical description of inclusive $\tau$~lepton hadronic decay is
performed in the framework of Dispersive approach to QCD. The significance
of effects due to hadronization is convincingly demonstrated. The approach
on hand is capable of describing experimental data on $\tau$~lepton decay
in vector and axial--vector channels. The vicinity of values of QCD scale
parameter~$\Lambda$ obtained in both channels testifies to the
self--consistency of developed approach.

\end{document}